\documentclass[prl,twocolumn,preprintnumbers,superscriptaddress,amsmath,amssymb]{revtex4-1}
\usepackage{graphicx}
\usepackage{subfigure}
\usepackage{mathrsfs}
\usepackage{amsfonts}
\usepackage{times}
\usepackage{amsmath}
\usepackage{leftidx}
\usepackage{tikz}
\usepackage{color}
\usepackage[colorlinks,linkcolor=blue,citecolor=blue]{hyperref}

\usepackage{bbold}
\usepackage{braket}
\usepackage{mathtools}

\begin{document}
\title{Thermodynamic Uncertainty Relation for Arbitrary Initial States}
\author{Kangqiao Liu}
\email{kqliu@cat.phys.s.u-tokyo.ac.jp}
\affiliation{Department of Physics, The University of Tokyo, 7-3-1 Hongo, Bunkyo-ku, Tokyo 113-0033, Japan}
\author{Zongping Gong}
\affiliation{Department of Physics, The University of Tokyo, 7-3-1 Hongo, Bunkyo-ku, Tokyo 113-0033, Japan}
\author{Masahito Ueda}
\affiliation{Department of Physics, The University of Tokyo, 7-3-1 Hongo, Bunkyo-ku, Tokyo 113-0033, Japan}
\affiliation{RIKEN Center for Emergent Matter Science, 2-1, Hirosawa, Wako-shi, Saitama, 351-0198, Japan}
\affiliation{Institute for Physics of Intelligence, The University of Tokyo, 7-3-1 Hongo, Bunkyo-ku, Tokyo 113-0033, Japan}
\date{\today}

\begin{abstract}
The thermodynamic uncertainty relation (TUR) describes a trade-off relation between nonequilibrium currents and entropy production and serves as a fundamental principle of nonequilibrium thermodynamics. However, currently known TURs presuppose either specific initial states or an infinite-time average, which severely limits the range of applicability. Here we derive a finite-time TUR valid for arbitrary initial states from the Cram\'er-Rao inequality. We find that the variance of an accumulated current is bounded by the instantaneous current at the final time, which suggests that ``the boundary is constrained by the bulk". We apply our results to feedback-controlled processes and successfully explain a recent experiment which reports a violation of a modified TUR with feedback control. We also derive a TUR that is linear in the total entropy production and valid for discrete-time Markov chains with non-steady initial states. The obtained bound exponentially improves the existing bounds in a discrete-time regime.
\end{abstract}
\maketitle


\emph{Introduction.---} Over the last two decades, stochastic thermodynamics \cite{Seifert2012,Sekimoto1998} has provided a general framework for understanding dissipation and thermal fluctuations far from equilibrium. 
Among the most important achievements are the fluctuation theorems \cite{Jarzynski1997,Crooks1999,Hatano2001,Seifert2005,Esposito2010,Sagawa2010,Sagawa2012,Gong2015}, which refine various second-law inequalities into equalities. Recently, yet another rigorous result known as the thermodynamic uncertainty relation (TUR) was discovered \cite{Seifert2015}, which dictates that the precision of a nonequilibrium time-integrated current observable $J$ be bounded from below by the inverse of the total entropy production (EP) $\sigma$:
\begin{equation}
\mathcal{Q}_{\rm C}\equiv\frac{\textrm{Var}[J]}{\langle J\rangle^2}\sigma\ge 2,
\label{CTUR}
\end{equation}
where $\langle J\rangle$ and $\textrm{Var}[J]$ are the average and the variance of $J$. The inequality (\ref{CTUR}) was originally discovered in biochemical networks \cite{Seifert2015} and proved by large deviation theory \cite{Gingrich2016}. 

The original TUR (\ref{CTUR}) has a rather limited range of applicability \cite{Horowitz2019}, where the system is assumed to obey a Markovian continuous-time dynamics and should start from a nonequilibrium steady state (NESS) \cite{Seifert2017,Horowitz2017} or wait until the system relaxes to the NESS \cite{Gingrich2016}. Without any one of these assumptions the bound could be violated \cite{Proesmans2017,Horowitz2019,Barato2018,Koyuk2018,vu2019uncertainty}. A number of generalizations have been discussed, such as discrete-time Markov chains \cite{Proesmans2017}, periodically driven systems \cite{Barato2018,Koyuk2018}, measurement and feedback control \cite{Potts2019,vu2019uncertainty}, active matter systems \cite{Shreshtha2019,Cao2019,Lee2019}, and quantum Markovian dynamics \cite{Carollo2019}. In particular, fluctuation theorems are found to directly lead to a bound involving an exponentiated EP, which is known as the generalized TUR (GTUR) \cite{Hasegawa2019prl,Timpanaro2019,Esposito2019}. Information-theoretic approaches such as the Martingale theory \cite{Pigolotti2017} and the Cram\'er-Rao inequality \cite{Dechant2018,Hasegawa2019pre,Vuarxiv2019,Hasegawaarxiv2019} have been utilized to derive the original TUR and its variants. 

\begin{figure}
\begin{center}
\includegraphics[width=8.5cm]{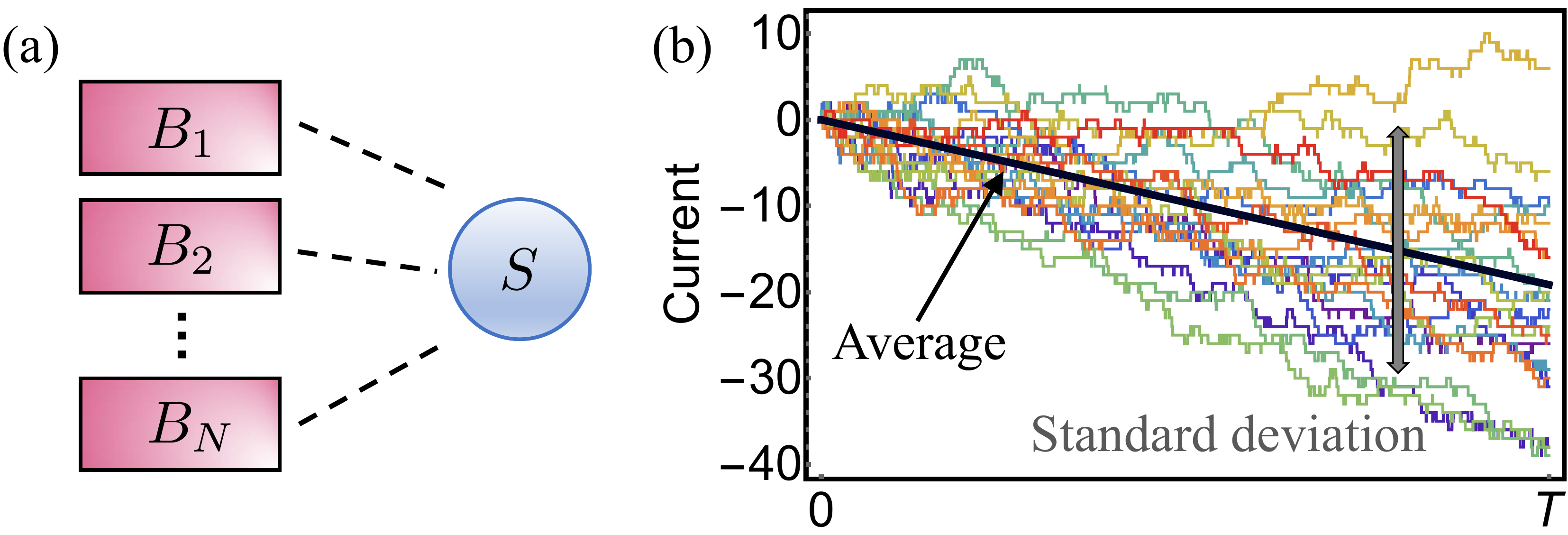}
\end{center}
\caption{(a) Schematic illustration of our setup. The system of interest $S$ is coupled with several thermal reservoirs $B_{\nu}$. The dynamics of the system is governed by a Markovian master equation (\ref{me}). The state transition at each time is caused by the reservoirs. The current can be any heat flow from the system to one of the reservoirs or a linear combination thereof. (b) Monte Carlo simulation of the time-integrated current in a two-heat-bath minimal model with a nonequilibrium steady state. Different colors correspond to different realizations. The black line indicates the average current $\langle J\rangle$. The length of the gray double arrow shows twice the standard deviation of the accumulated current.}
\label{fig1}
\end{figure}

However, none of these generalizations are quite satisfactory because their bounds are either very loose such as the GTURs or involving terms with no clear physical meaning. Moreover, most of these bounds require an initialization to a NESS or other specific states. 
In this Letter, we fill the gaps by deriving universal bounds on fluctuation and dissipation valid for an arbitrary finite time and arbitrary initial states in continuous-time and discrete-time Markov processes via the Cram\'er-Rao inequality. For continuous-time processes, our bound is a highly nontrivial generalization of Eq.~(\ref{CTUR}), where the ensemble-averaged time-integrated current $\langle J\rangle$ is replaced by the final-time instantaneous current multiplied by the time period, which implies that the boundary current is constrained by the bulk fluctuation and EP. Our formula reduces to the original TUR when the initial state is a NESS. We illustrate our result with minimal models and apply it to feedback-controlled processes. In particular, we explain a recent experiment which reports a violation in a modified TUR with feedback control \cite{Paneru2019}. For discrete-time processes, we find that the total EP modified by a certain sum of Kullback-Leibler divergences should be rescaled by the minimal staying probability of the Markov chain. Our result exponentially improves the existing results in a discrete-time regime \cite{Proesmans2017,Pigolotti2018}.


\emph{Setup.---}  We consider a general multi-channel Markovian system $S$ (see Fig.~\ref{fig1}) described by the master equation
\begin{equation}
\dot{\mathbf{P}}(t)=\mathcal{R}\mathbf{P}(t),
\label{me}
\end{equation}
where $[\mathbf{P}(t)]_x\equiv P(x;t)$ is the system-state distribution at time $t$ and $[\mathcal{R}]_{yx}\equiv r(x,y)
=\sum_{\nu}r^{\nu}(x,y)$ is the time-independent transition rate matrix. Here $r^{\nu}(x,y)$ is the transition rate from $x$ to $y$ via channel $\nu$, i.e., the transition is induced by the $\nu$th heat bath $B_\nu$ at inverse temperature $\beta_\nu$. For a trajectory $\omega=(x_0, t_0=0; x_1,\nu_1, t_1; x_2,\nu_2,t_2;\dots; x_n,\nu_n, t_n\le T\equiv t_{n+1})$, where a transition from $x_{j-1}$ to $x_j$ via channel $\nu_j$ occurs at $t_j$ ($j=1,2,...,n$) during a finite time period $T$, the path probability density governed by the master equation (\ref{me}) is given by 
\begin{equation}
\mathcal{P}[\omega]=P(x_0)e^{-\sum_{x}\lambda(x)\tau_x[\omega]+\sum_{x\ne y,\nu}n^{\nu}_{xy}[\omega]\ln r^{\nu}(x,y)},
\label{PP}
\end{equation}
where $P(x_0)$ is the initial distribution, $\lambda(x)=\sum_{y: y\ne x}r(x,y)$ is the escape rate for state $x$,  $\tau_x[\omega]\equiv\sum^n_{j=0}\delta_{x_jx}(t_{j+1}-t_j)$ is the total time during which the system stays in state $x$, and $n^{\nu}_{xy}[\omega]\equiv\sum^n_{j=1}\delta_{x_{j-1}x}\delta_{x_jy}\delta_{\nu_j\nu}$ is the total number of transitions from $x$ to $y$ through channel $\nu$. A general stochastic accumulated current is defined as
\begin{equation}
J[\omega]\equiv \sum_{x\ne y,\nu}n^{\nu}_{xy}[\omega]d^{\nu}(x,y),
\label{J}
\end{equation}
where $d^{\nu}(x,y)=-d^{\nu}(y,x)$ is the anti-symmetric increment associated with transition $x\to y$ via channel $\nu$. For example, $d^\nu(x,y)=\delta_{x_0x}\delta_{y_0y}\delta_{\nu_0\nu}$ gives the net number of transitions from $x_0$ to $y_0$ via channel $\nu_0$, while $d^\nu(x,y)=(E_x-E_y)\delta_{\nu_0\nu}$ ($E_x$: energy of state $x$) gives the net heat flow into the $\nu_0$th bath. Provided that the local detailed balance $r^\nu(x,y)e^{-\beta_\nu E_x}=r^\nu(y,x)e^{-\beta_\nu E_y}$ holds, the ensemble-averaged total EP for the dynamics is given by \cite{Esposito2010pre}
\begin{equation}
\sigma=\int_{0}^{T}dt\sum_{x\ne y,\nu}P(x;t)r^{\nu}(x,y)\ln\frac{P(x;t)r^{\nu}(x,y)}{P(y;t)r^{\nu}(y,x)}.
\label{EP}
\end{equation}

\emph{Main result.---} We show that the fluctuation of an arbitrary accumulated current (\ref{J}) is bounded by the total EP and the final instantaneous current:
\begin{equation}
\mathcal{Q}_{\rm T}\equiv\frac{\textrm{Var}[J]}{(Tj(T))^2}\sigma\ge 2,
\label{TTUR}
\end{equation}
where $j(T)\equiv \sum_{x\ne y,\nu}P(x; T)r^{\nu}(x,y)d^{\nu}(x,y)$ is the ensemble-averaged final instantaneous current. Such an inequality can equivalently be written as
\begin{equation}
|j(T)|\le\sqrt{\frac{1}{2}\textrm{Var}\left[\frac{J}{T}\right]\sigma}, 
\label{bbb}
\end{equation}
which implies that the boundary current is constrained by the bulk (time-averaged) current fluctuation and dissipation. In other words, if we want to achieve a large instantaneous current, which means driving the system far from equilibrium, we should either suffer large dissipation or sacrifice the quality (small fluctuation) of the time-integrated current. This statement refines the dissipation-precision trade-off of conventional TURs for NESSs \cite{Seifert2015,Gingrich2016,Seifert2017,Horowitz2017}.

Some remarks are in order here. First, the bound (\ref{TTUR}) holds for arbitrary initial states and there can be multiple transition channels. If there is only a single heat bath and the initial state is a NESS, the denominator is nothing but the accumulated current; thus the original TUR (\ref{CTUR}) is recovered. Second, every term in our bound allows a clear physical interpretation and is experimentally measurable \cite{Schuler2005,Koski2013,Koski2014}.  
Previous efforts at generalizing the TUR mainly focus on modifying the EP \cite{Barato2018,Koyuk2018,Vuarxiv2019,Hasegawaarxiv2019}; however, the modification lacks a clear physical meaning. Third, our result implies a sufficient (necessary) condition for the validity (violation) of the original TUR (\ref{CTUR}) for a non-steady initial state -- the final current $j(T)$ is larger (smaller) than the time-averaged one $\bar j\equiv\langle J\rangle/T$. This is primarily due to an increase (decrease) of the current, as we will illustrate in some minimal models. Finally, we emphasize that even the widely applicable GTUR generally breaks down for an arbitrary initial state since it requires that the initial and final states coincide \cite{Hasegawa2019prl}.

\begin{figure}
\begin{center}
\includegraphics[width=8.5cm]{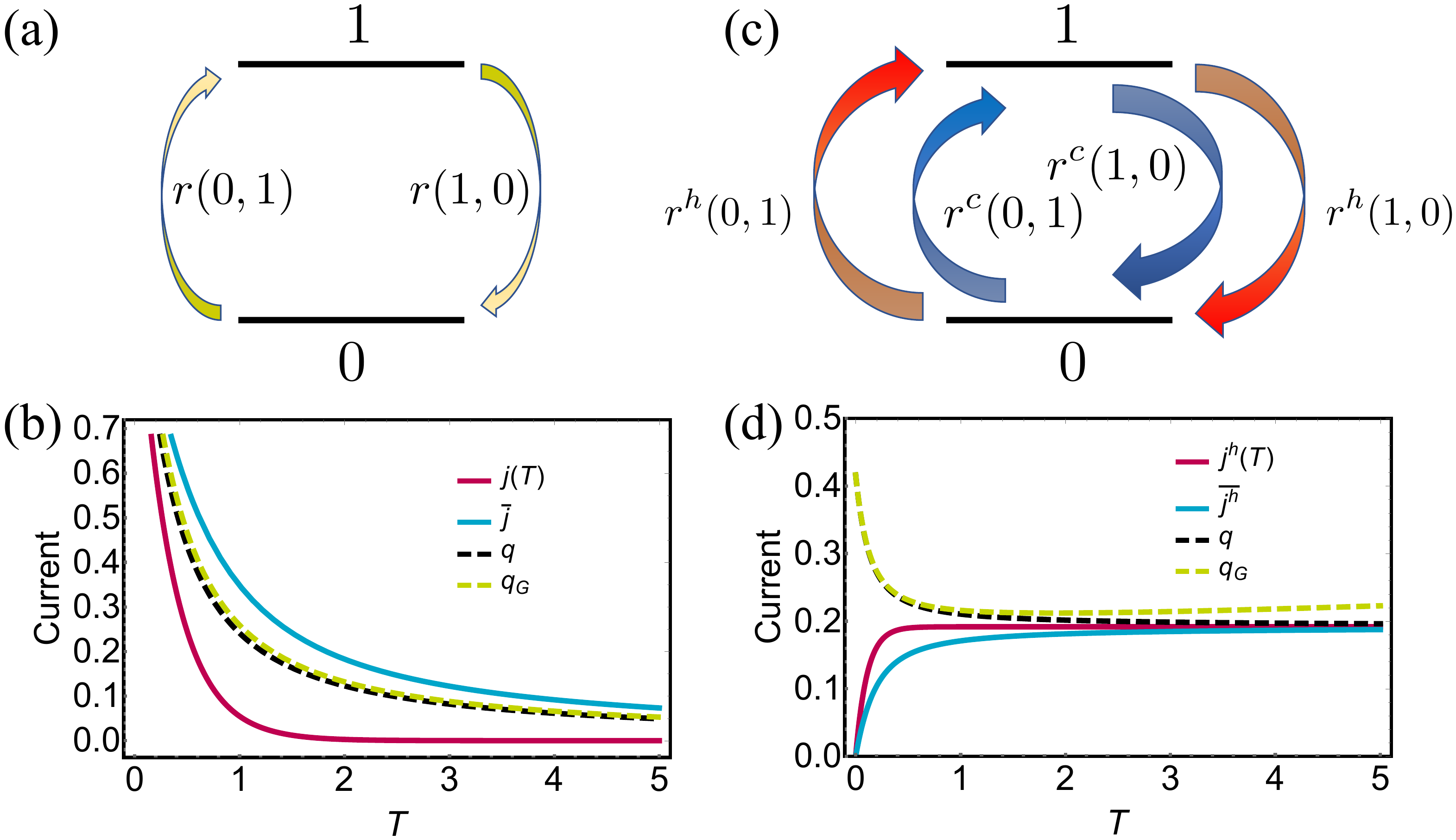}
\end{center}
\caption{(a) Two-level system coupled with a single heat bath. (b) Comparison between the final instantaneous current (red), the time-averaged current $\bar j\equiv\langle J\rangle/T$ (blue) and the current bounds from the conventional (black dashed, $q\equiv T^{-1}\sqrt{\textrm{Var}[J]\sigma/2}$) and generalized (green dashed, $q_{G}\equiv T^{-1}\sqrt{\textrm{Var}[J](e^{\sigma}-1)/2}$) TURs. Our inequality (\ref{bbb}) is satisfied, whereas the original one (\ref{CTUR}) is not. The initial state is chosen to be $\mathbf{P}(0)=[0.3,0.7]^{\mathbb{T}}$ and the transition rates are $r(0,1)=1$ and $r(1,0)=2$. (c) Two-level system coupled with cold and hot baths. Red arrows represent the transitions by coupling with the hot heat bath, and blue ones represent the cold bath. (d) Same quantities as in (b) for the model in (c), where both inequalities (\ref{TTUR}) and (\ref{CTUR}) are valid. The initial state is chosen so that the initial hot current vanishes. We set $\beta_h=1$, $\beta_c=1.5$, $r^{h}(0,1)=r^{c}(0,1)=1$ in our simulation.}
\label{fig2}
\end{figure}


\emph{Two minimal models.---} Before going into the derivation of the main result, let us examine the main result (\ref{TTUR}) in some minimal models. We first consider the simplest example for an equilibrium steady state. As shown in Fig.~\ref{fig2}(a), a two-level system with states $0$ and $1$ couples to a single heat reservoir at inverse temperature $\beta$. The energy gap between the two states is set to be $\Delta=1$ and the state $0$ is assumed to be lower in energy. We start from an arbitrary initial state $\mathbf{P}(0)=[p,1-p]^{\mathbb{T}}$ ($\mathbb{T}$: transpose) and let the system relax to its equilibrium steady state. The current is chosen to be the net flow from $1$ to $0$. According to the local detailed balance condition, two transition rates $r(0,1)$ and $r(1,0)$ satisfy $r(0,1)=e^{-\beta}r(1,0)$. By utilizing full counting statistics \cite{deJong1996,Bagrets2003,Ren2010}, we can analytically calculate all the quantities in the bound (\ref{TTUR}) \cite{SM}. In Fig.~\ref{fig2}(b), we find that only our bound holds with the conventional TUR and the GTUR being violated. This is because the currents decrease exponentially with time, implying that the time-averaged current is larger than the final time current. Consequently, our $\mathcal{Q}_{\rm T}$ value should be larger than the conventional $\mathcal{Q}_{\rm C}$, and therefore the conventional TUR may fail.

We now consider a simplest model for an NESS which involves a two-level system with states $0$ and $1$ and the energy gap $\Delta=1$ in contact with two heat baths at inverse temperatures $\beta_h$ and $\beta_c$. Since the state transition can be induced by either of the baths, there is a total of 4 transition rates (see Fig.~\ref{fig2}(c)) which satisfy two local detailed balance relations: $r^{h}(1,0)=e^{\beta_h}r^{h}(0,1)$ and $r^{c}(1,0)=e^{\beta_c}r^{c}(0,1)$. The current is chosen to be the heat flow from the hot bath, whose instantaneous value at time $t$ reads $j^{h}(t)=P_0(t)r^{h}(0,1)-P_1(t)r^{h}(1,0)$. We start from a special initial state so that $j^{h}(0)$ vanishes. The current fluctuation is again calculated from full counting statistics \cite{SM}. In Fig.~\ref{fig2}(d), we see that both $j^{h}(T)$ and $\overline{j^{h}}$ are bounded from above by $q$, while $j^{h}(T)$ is tighter. This is because the current monotonically increases so that the final current is larger than the time-averaged one. Accordingly, our $\mathcal{Q}_{\rm T}$ should be smaller than $\mathcal{Q}_{\rm C}$. Since $\mathcal{Q}_{\rm T}$ is bounded from below by $2$, the larger quantity $\mathcal{Q}_{\rm C}$ should be bounded from below by $2$ as well.

\begin{figure}
\begin{center}
\includegraphics[width=4.5cm]{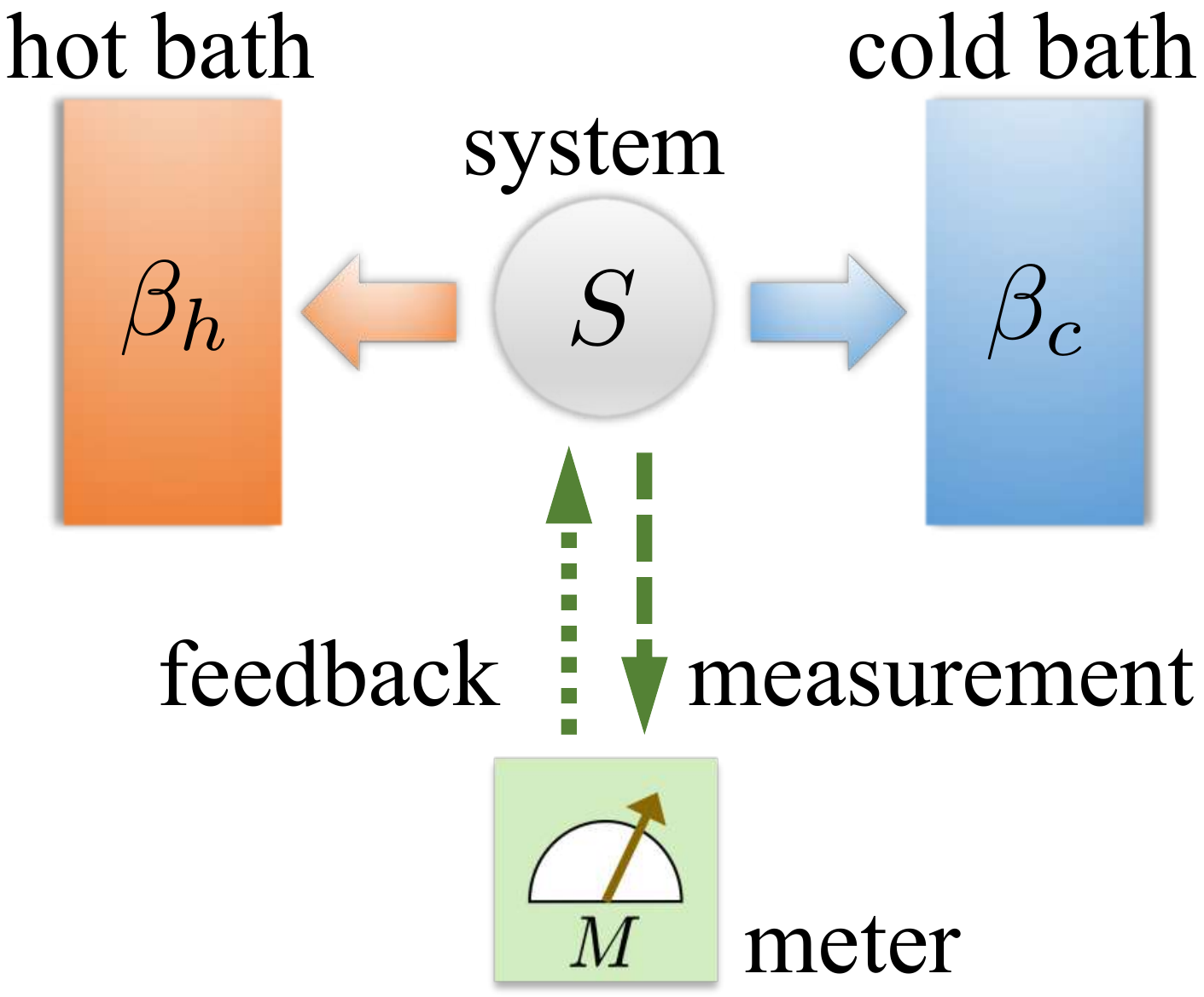}
\end{center}
\caption{System in contact with cold and hot baths subject to measurement and feedback. The system is probed by a meter $M$ and subject to feedback control according to the measurement outcome. After the feedback, the system undergoes Markovian dynamics by interacting with two heat baths. 
}
\label{fig3}
\end{figure}

\emph{Application to feedback-controlled processes.---} Our TUR (\ref{TTUR}) can readily be extended to include the effect of measurement and feedback control in the context of information thermodynamics \cite{Sagawa2015}. As a general setup, the system of interest $S$ couples with multiple heat baths $B_\nu$'s at inverse temperature $\beta^\nu$'s. In addition, as shown in Fig.~\ref{fig3}, a meter $M$ probes the state of the system and performs feedback control by adjusting the transition matrix into $\mathcal{R}_m$ according to the measurement outcome $m$ \cite{Sagawa2012pre}. We assume that the measurement and feedback are done instantaneously, after which the system will relax during a time interval $\tau$ through coupling to the baths. This assumption is justified if they are performed sufficiently fast compared with the stochastic transitions in the system \cite{Sagawa2012,Sagawa2012pre,Paneru2019}. At the end of the relaxation the meter will be reset to its default state, e.g. $0$, and then the next cycle begins \cite{Shiraishi2015}. The system will eventually reach a stroboscopic steady state, which is a periodic steady state in the sense that the state of the system will be statistically the same after one period while it can change during each cycle.  Within a single relaxation period, the total EP should be $\sigma=\mathcal{I}(0)-\mathcal{I}(\tau)+\sigma_{\rm S}+\sigma_{\rm B}$, where $\mathcal{I}(t)\equiv\sum_{s,m}P(s,m;t)\ln\frac{P(s,m;t)}{P_{\rm S}(s;t)P_{\rm M}(m)}$ is the mutual information between the system and the meter at time $t$ with $P(s,m;t)$, $P_{\rm S}(s;t)$ and $P_{\rm M}(m)$ being the joint distribution of the system and the meter, the marginal distribution of the system and that of the meter; $\sigma_{\rm S}$ and $\sigma_{\rm B}$ are the entropy changes in the system and baths, respectively. Here we have used the fact that $P_{\rm M}(m)$ is time-independent and thus there is no entropy production in the meter. Defining the consumed mutual information $\Delta\mathcal{I}\equiv\mathcal{I}(0)-\mathcal{I}(\tau)$ and the physical entropy production $\sigma_{\rm P}\equiv\sigma_{\rm S}+\sigma_{\rm B}$, we have
\begin{equation}
\mathcal{Q}_{\rm T}=\frac{{\rm Var}[J]}{(\tau j(\tau))^2}(\sigma_{\rm P}+\Delta\mathcal{I})\ge2,
\label{fbTUR}
\end{equation}
where $J$ can be an arbitrary current determined from an anti-symmetric increment $d^\nu_m(x,y)$ that generally depends on the measurement outcome. Note that if the system reaches a stroboscopic steady state, then $\sigma_{\rm P}=\sigma_{\rm B}$.

We can explain a recent experiment on feedback control \cite{Paneru2019} with the criteria described in the previous section. The authors in Ref.~\cite{Paneru2019} constructed an information engine consisting of an optically trapped colloidal particle immersed in a heat reservoir at inverse temperature $\beta$, following a repeated protocol of measurement, feedback and relaxation. In the $i$th cycle, the demon measures the position $x_i$ of the particle. Due to noise, the outcome $y_i$ could be different from $x_i$. The center of the potential $\lambda_{i-1}$ is suddenly shifted to $y_i$ and let the particle relax for a period $\tau$ before the next cycle begins, obeying the overdamped Langevin equation. The system will reach a stroboscopic steady state after many cycles. The stochastic current is the work $\beta W$ performed on the particle by shifting the potential. Because there is only one heat bath, the dynamics after feedback control is simply a relaxation process toward equilibrium. The absolute value of the current always decreases with time. The conventional TUR can be violated for a certain range of parameters as reported in Ref.~\cite{Paneru2019}.

\emph{Generalization to discrete-time Markov chains.---} We consider a general multi-channel Markovian system $S$ starting from an arbitrary initial state as in Fig.~\ref{fig1}(a) which is now described by the following discrete-time evolution equation:
\begin{align}
P(x,t_{i})=\sum_{y,\nu}A^{\nu}(x|y)P(y,t_{i-1}),\label{discreteME}
\end{align}
where $P(x,t_{i})$ is the probability of the system being in state $x$ at time $t_{i}$ and $A^{\nu}(x|y)$ is the transition probability from state $y$ to state $x$ through channel $\nu$. The transition probabilities satisfy the normalization condition: $\sum_{x,\nu}A^{\nu}(x|y) =1$. The total EP for $n$ steps is given by \cite{Morimoto1963,Lee2018,SM}
\begin{align}
\sigma=\sum_{i=1}^{n}\sum_{x,y,\nu}P(x,t_{i-1})A^{\nu}(y|x)\ln\frac{P(x,t_{i-1})A^{\nu}(y|x)}{P(y,t_i)A^{\nu}(x|y)}.\label{discreteEP}
\end{align}
The TUR valid for this process \eqref{discreteME} reads \cite{SM}
\begin{align}
\mathcal{Q}_{\rm D}\equiv\frac{\textrm{Var}[J]}{(nj(t_{n-1}))^2}\frac{\tilde{\sigma}}{a}\ge2,\label{NSDT}
\end{align}
where the tilde EP is defined as $\tilde{\sigma}\equiv\sigma+\sum_{i=1}^{n}D_{\textrm{KL}}(\mathbf{P}(t_i)||\mathbf{P}(t_{i-1}))$ with $D_{\rm KL}$ being the Kullback-Leibler divergence, $a$ is the minimal staying probability $a\equiv\min_{x}A(x|x)$, and $j(t_{n-1})\equiv \sum_{x\ne y,\nu}P(x,t_{n-1})A^{\nu}(y|x)d^{\nu}(y|x)$ is the current at the final step.  We make two comments. First, the bound \eqref{NSDT} can be reduced to the continuous-time bound \eqref{TTUR} in the limit of $\Delta t\to 0$. Second, there exists a discrete-time TUR exponentiated in the total EP for NESS \cite{Proesmans2017}. Our bound \eqref{NSDT} exponentially improves the result because it is linear in the total EP.

\emph{Derivation of the main result.---} We finally prove inequality (\ref{TTUR}). We can employ large deviation theory to derive our result (\ref{TTUR}) \cite{SM,PietzonkaPhD}. However, a more straightforward and elegant approach is based on the generalized Cram\'er-Rao inequality \cite{Cover2006}:
\begin{equation}
\textrm{Var}_{\theta}\big[\Theta\big]\ge \frac{\psi'(\theta)^2}{F(\theta)},
\label{CR}
\end{equation}
where $\theta$ is a parameter, $F(\theta)$ is the Fisher information and $\Theta[\omega]$ is an unbiased estimator for a smooth function $\psi(\theta)$, i.e., $\langle \Theta \rangle_{\theta}=\psi(\theta)$. Here, the average is defined as $\langle g \rangle_{\theta}\equiv \int \mathcal{D}\omega g[\omega] \mathcal{P}_{\theta}[\omega]$ for a parametrized distribution $\mathcal{P}_{\theta}[\omega]$. 
Our goal is to relate each term in (\ref{CR}) to the thermodynamic quantities in Eq.~(\ref{TTUR}) \cite{Hasegawa2019pre,Vuarxiv2019,Hasegawaarxiv2019}. To this end, we first parametrize a typical path probability density in Eq.~(\ref{PP}) as 
\begin{equation}
\mathcal{P}_{\theta}[\omega]=P_{\theta}(x_0)e^{\sum^n_{j=1}\ln r_{\theta}^{\nu_j}(x_{j-1},x_j;t_j)-\sum^n_{j=0}\int^{t_{j+1}}_{t_j}dt\lambda_{\theta}(x_j;t)},
\label{PPtheta}
\end{equation}
which is determined from an auxiliary transition matrix $\mathcal{R}_\theta(t)$ with time-dependent entries $[\mathcal{R}_\theta(t)]_{yx}=r_{\theta}^{\nu}(x,y;t)$ and $[\mathcal{R}_\theta(t)]_{xx}=-\lambda_\theta(x;t)\equiv-\sum_{y:y\neq x,\nu}r_{\theta}^{\nu}(x,y;t)$. When we set $\theta$ to be a certain value, say $0$, $r_{\theta}^{\nu}(x,y;t)$ should go back to the time-independent typical value $r^{\nu}(x,y)$. By definition, the Fisher information can be calculated from Eq.~(\ref{PPtheta}) as
\begin{equation}
F(\theta)=\int_{0}^{T}dt\sum_{x\ne y,\nu}P_{\theta}(x;t)r^{\nu}_{\theta}(x,y;t)\bigg(\frac{\partial}{\partial \theta}\ln r^{\nu}_{\theta}(x,y;t)\bigg)^2,
\label{FI}
\end{equation}
where the initial auxiliary state has been assumed to be a typical one, i.e., $P_{\theta}(x_0)=P(x_0)$.
By choosing $\Theta[\omega]=\sum^n_{j=1}d^{\nu_j}(x_{j-1},x_j)$ to be a general accumulated current, ${\rm Var}_\theta[\Theta]$ at $\theta=0$ simply gives the desired current fluctuation. In this case, $\psi(\theta)$ is nothing but the ensemble-averaged current given by
\begin{eqnarray}
\langle \Theta\rangle_{\theta}=\int_{0}^{T}dt\sum_{x\ne y,\nu}P_{\theta}(x;t)r^{\nu}_{\theta}(x,y;t)d^{\nu}(x,y), 
\label{Tth}
\end{eqnarray}
where $P_{\theta}(x;t)$ is determined from the solution of the parameterized master equation with generator $\mathcal{R}_\theta(t)$.

Comparing the structure of our TUR (\ref{TTUR}) with the Cram\'er-Rao inequality (\ref{CR}), we relate the Fisher information $F(0)$ and $\psi'(0)$ to half of the total EP $\sigma$ given in Eq.~(\ref{EP}) and the final current $j(T)$, respectively. As a sufficient condition, we choose the parametrization $r_\theta^\nu(x,y;t)=r^\nu(x,y)e^{\theta\alpha^\nu_{xy}(t)}$, 
and assume that for each pair of $(x,y)$ at any time $t$ and any channel $\nu$, the following conditions are satisfied:
\begin{eqnarray}
&K^\nu_{xy}(\alpha^\nu_{xy})^2+K^\nu_{yx}(\alpha^\nu_{yx})^2=\frac{1}{2}(K^\nu_{xy}-K^\nu_{yx})\ln\frac{K^\nu_{xy}}{K^\nu_{yx}}, \\
&K^\nu_{xy}\alpha^\nu_{xy}-K^\nu_{yx}\alpha^\nu_{yx}=K^\nu_{xy}-K^\nu_{yx}, \label{Keq}
\end{eqnarray}
where $K^\nu_{xy}(t)\equiv P(x;t)r^{\nu}(x,y)$ and its time dependence (as well as that in $\alpha^\nu_{xy}(t)$) is omitted for simplicity. The above two equations, whose solutions always exist \cite{SM}, guarantee $F(0)=\frac{1}{2}\sigma$ and $\psi'(0)=Tj(T)$ for an arbitrary $d^\nu(x,y)$. The former simply follows from Eqs.~(\ref{EP}) and (\ref{FI}). To show the latter, we note that, up to the leading (first) order in $\theta$, the parametrized probability is given by \cite{SM}
 \begin{equation}
 \mathbf{P}_\theta(t)=  \mathbf{P}(t)+\theta t \dot{\mathbf{P}}(t)+O(\theta^2),
 \end{equation}
leading to  $\partial_{\theta}\mathbf{P}_\theta(t)|_{\theta=0}= t \dot{\mathbf{P}}(t)$. Combining this result with Eq.~(\ref{Keq}), we find that $\psi'(0)\equiv\partial_\theta\langle\Theta\rangle|_{\theta=0}$ is an integral of a total derivative $\frac{d}{dt}[t j(t)]$ with $j(t)\equiv \sum_{x\neq y,\nu}P(x;t)r^\nu(x,y)d^\nu(x,y)$ being the instantaneous current. Therefore, we obtain $\psi'(0)=tj(t)|^T_0=Tj(T)$. For the case with feedback control, we have only to add another index $m$ representing the meter's state.

\emph{Summary and outlook.---} We have established new TURs (\ref{TTUR}) and \eqref{NSDT} for general continuous- and discrete-time multi-channel Markovian systems starting from an arbitrary initial state. Our results includes the conventional TURs \cite{Seifert2015,Gingrich2016,Seifert2017,Horowitz2017} as special cases and incorporate the effect of measurement and feedback control (see inequality~(\ref{fbTUR})). The continuous bound (\ref{TTUR}) can also be used to explain the recent experiment \cite{Paneru2019}. The discrete bound \eqref{NSDT} exponentially improves the TURs in a discrete-time regime. While our results greatly extend the range of validity of the TURs, the time-homogeneous assumption of transition rates and probabilities needs to be made. How to relax this requirement is an important subject for future studies. It should also be of interest to investigate the effect of absolute irreversibility \cite{Murashita2014,Murashita2017} on the TURs.

We acknowledge Chikara Furusawa, Yuto Ashida, David H. Wolpert, Takahiro Sagawa and Shun Otsubo for valuable discussions. K.L. was supported by Global Science Graduate Course (GSGC)
program of the University of Tokyo. Z.G. was supported by MEXT. This work was supported
by KAKENHI Grant No. JP18H01145 and a Grant-in-Aid
for Scientific Research on Innovative Areas “Topological
Materials Science (KAKENHI Grant No. JP15H05855) from
the Japan Society for the Promotion of Science.
\bibliography{kqliu_references_v2}

\clearpage
\begin{center}
\textbf{\large Supplemental Material}
\end{center}
\setcounter{equation}{0}
\setcounter{figure}{0}
\makeatletter
\renewcommand{\theequation}{S\arabic{equation}}
\renewcommand{\thefigure}{S\arabic{figure}}
\renewcommand{\bibnumfmt}[1]{[S#1]}

\section{Minimal model for equilibrium steady states}
In our model, the initial state is $\mathbf{P}(0)=[p,1-p]^{\mathbb{T}}$ and two transition rates satisfy the local detailed balance condition $r(0,1)=e^{-\beta}r(1,0)$. For simplicity, we denote $r(0,1)=a$ and $r(1,0)=b$. Then the transition rate matrix reads
\begin{align}
\mathcal{R}=\left[ \begin{array}{cc}
-a & b \\
a & -b
\end{array} \right].
\end{align}
In terms of the transition rate matrix, the final state can be related to the initial state as
\begin{align}
\mathbf{P}(T)=e^{\mathcal{R}T}\mathbf{P}(0)=[P_0(T),P_1(T)]^{\mathbb{T}},
\end{align}
where
\begin{align}
P_0(T)=\frac{e^{-(a+b)T}}{a+b}[ap+b(-1+p+e^{(a+b)T})]
\end{align}
and $P_1(T)=1-P_0(T)$.
The ensemble-averaged accumulated current $\langle J \rangle$ and its variance $\textrm{Var}[J]$ can be calculated by means of full counting statistics \cite{deJong1996,Bagrets2003,Ren2010}. By introducing a counting field $\chi$ in the rate matrix, the characteristic function should read
\begin{align}
\mathcal{Z}(\chi)=[1,1]e^{\mathcal{R}(\chi)T}\mathbf{P}(0),
\end{align}
where
\begin{align}
\mathcal{R}(\chi)=\left[ \begin{array}{cc}
-a & be^{-i\chi} \\
ae^{i\chi} & -b
\end{array} \right].
\end{align}
After a straightfoward calculation, the characteristic function is given by
\begin{align}
\mathcal{Z}(\chi)=&\frac{1}{a+b}\big[(b+ae^{-(a+b)T})p+(a+be^{-(a+b)T})(1-p)\nonumber\\
&+be^{-(a+b)T}e^{-i\chi}(-1+e^{(a+b)T})(1-p)\nonumber\\
&+ae^{-(a+b)T}e^{i\chi}(-1+e^{(a+b)T})p\big].
\end{align}

\begin{figure}
\begin{center}
\includegraphics[width=8.5cm]{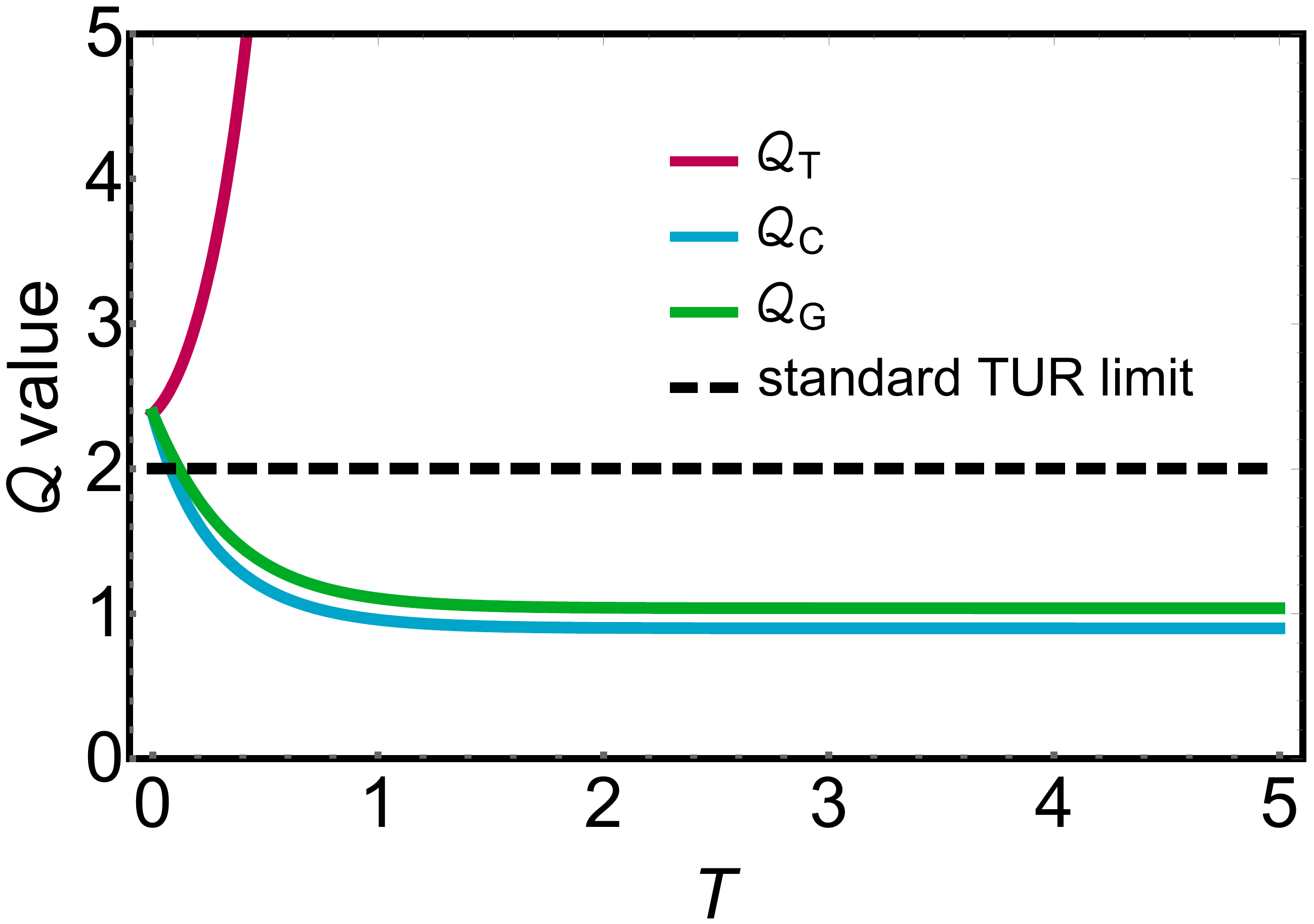}
\end{center}
\caption{$\mathcal{Q}$ values defined in the main text versus time $T$. The red solid curve is $\mathcal{Q}_{\rm T}$ in our TUR. The blue and green curves show the $\mathcal{Q}$ values of the conventional TUR and the GTUR, respectively. As in Fig.~2(b) in the main text, our inequality always holds, whereas the original TUR and the GTUR are violated.}
\label{figS1}
\end{figure}

The mean and variance of the current can be obtained from the first and second derivatives of $\ln\mathcal{Z}(\chi)$ with respect to $i\chi$ at $\chi=0$ as
\begin{align}
\langle J\rangle=\frac{b (-1 + p) + a p}{a + b}(1 - e^{-(a + b) T}),
\end{align}
\begin{widetext}
\begin{equation}
\textrm{Var}[J]=
\frac{e^{-2 (a + b) T}}{(a + b)^2}\big[-[b (-1 + p) + a p]^2 + 
   e^{2 (a + b) T} [a b + (a + b)^2 p - (a + b)^2 p^2] + 
   e^{(a + b) T} [b (-a + b) - (a + b) (a + 3 b) p + 
      2 (a + b)^2 p^2]\big].
\end{equation}
\end{widetext}
The total entropy production $\sigma$ involves two contributions. One is the Shannon entropy change in the system:
\begin{align}
\sigma_{\rm S}=S(T)-S(0),
\end{align}
where $S(t)=-[P_0(t)\ln P_0(t)+P_1(t)\ln P_1(t)]$. The other is the entropy production in the heat bath $\sigma_{\rm B}=-\beta \langle J\rangle$. Thus all terms involved in the TUR are analytically obtained. In the simulation, we set $p=0.3$, $r(0,1)=1$ and $r(1,0)=2$.

In Fig.~\ref{figS1}, the $Q$ values of our inequality, the original TUR and the GTUR are presented. We find that while the conventional TUR and the GTUR break down, our bound remains valid, which is consistent with Fig.~2(b) in the main text.

\section{Minimal model for nonequilibrium steady states}
In this model, the only modification from the previous example is that there are 4 transition rates satisfying 2 local detailed balance conditions: $r^{h}(1,0)=e^{\beta_h}r^{h}(0,1)$ and $r^{c}(1,0)=e^{\beta_c}r^{c}(0,1)$. We introduce the following short-hand notations: $r^{h}(0,1)=a_h$, $r^{h}(1,0)=b_h$, $r^{c}(0,1)=a_c$ and $r^{c}(1,0)=b_c$. The rate matrix is then given by
\begin{align}
\mathcal{R}(\chi)=\left[ 
\begin{array}{cc} -a_c-a_h&b_c+b_he^{-i\chi}\\ a_c+a_he^{i\chi}&-b_c-b_h\end{array} \right].
\end{align}
The instantaneous hot current is $j^{h}(t)=P_0(t)r^{h}(0,1)-P_1(t)r^{h}(1,0)$. We start from an initial state where $j^{h}(0)$ vanishes, i.e.,
\begin{align}
\mathbf{P}(0)=\bigg[\frac{1}{1+e^{-\beta_h},}\frac{e^{-\beta_h}}{1+e^{-\beta_h}}\bigg]^{\mathbb{T}}\equiv [\alpha, \beta]^{\mathbb{T}}.
\end{align}

The characteristic function can explicitly be calculated in terms of
\begin{align}
D(\chi)=\sqrt{4mn+(a-b)^2},
\end{align}
with $a=-a_c-a_h$, $m(\chi)=b_c+b_he^{-i\chi}$, $n(\chi)=a_c+a_he^{i\chi}$ and $b=-b_c-b_h$, as
\begin{align}
\mathcal{Z}(\chi)&=e^{\frac{a+b}{2}\tau}\big\{(\alpha+\beta)\cosh(\frac{D}{2}\tau)\nonumber\\&+\frac{1}{D}[(a-b+2n)\alpha-(a-b-2m)\beta]\sinh(\frac{D}{2}\tau)\big\}.
\label{Z}
\end{align}
Similarly, the first and second derivatives of the characteristic function give analytic expressions of the mean and the variance. The entropy production of the bath is given by $\sigma_{\rm B}=\beta_h \langle J^h\rangle+\beta_c \langle J^c\rangle$, where the mean cold and hot currents can be calculated similarly to the single-bath case (see Sec.~I). We set $\beta_h=1$, $\beta_c=1.5$, $r^{h}(0,1)=r^{c}(0,1)=1$ in our simulation.

\begin{figure}
\begin{center}
\includegraphics[width=8.5cm]{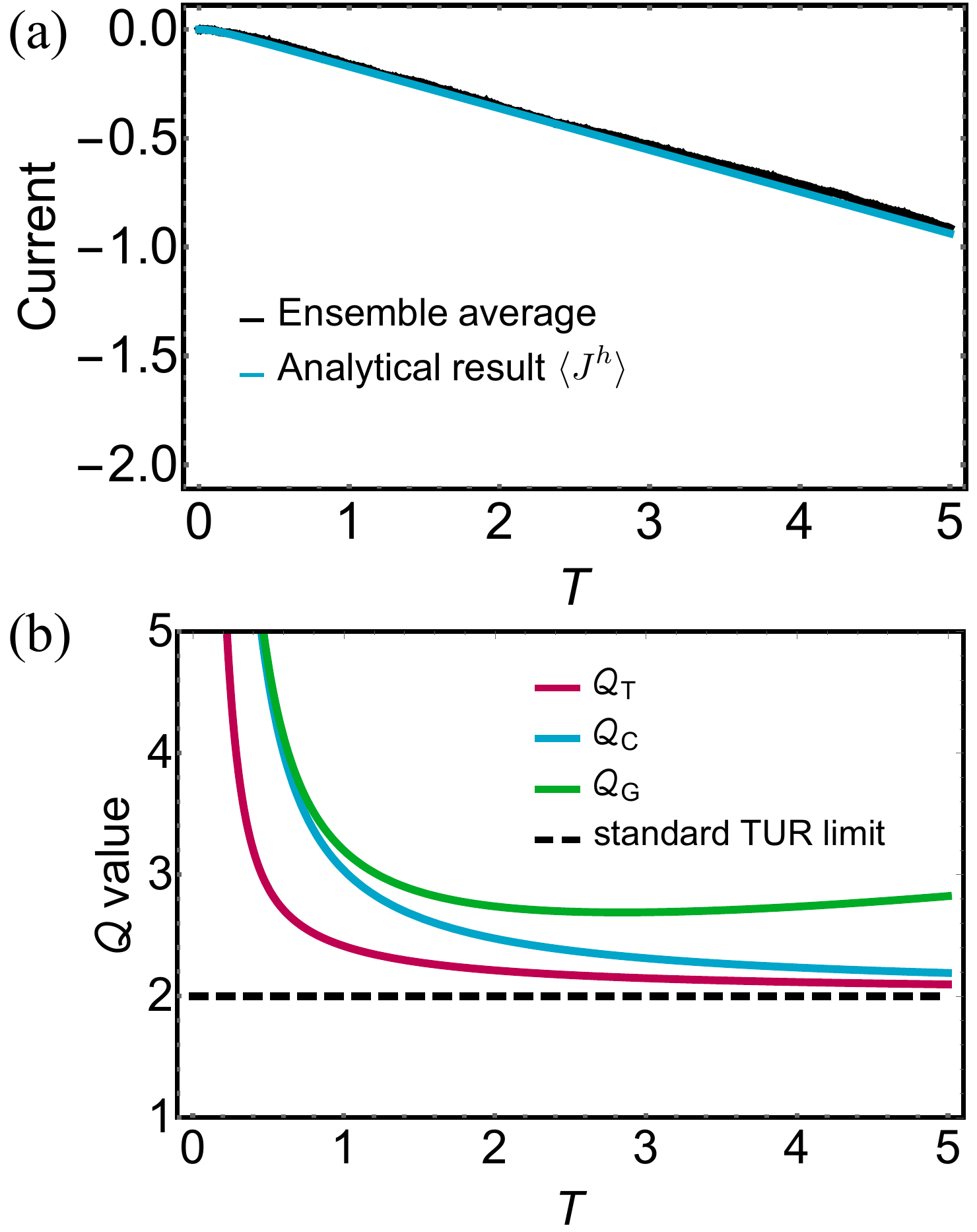}
\end{center}
\caption{(a) Time dependence of the current obtained from the average over $10000$ trajectories of Monte Carlo simulations. Superimposed is the analytic result calculated from full counting statistics. An excellent agreement confirms the correctness of our calculation. (b) $\mathcal{Q}$ values for the conventional TUR (blue), the GTUR (green) and our TUR (red). We find that while all TURs are valid, ours is the tightest, which is consistent with Fig.~2(d) in the main text.}
\label{figS2}
\end{figure}

In Fig.~\ref{figS2}, we see that while all TURs are satisfied, ours is tightest, which is consistent with Fig.~2(d) in the main text.

\section{Application to feedback control}
Let us verify inequality (8) in the main text for the two-level minimal model. The only difference from the minimal model for an NESS is that there is a two-state meter that probes the system and performs feedback control. As a result of two control protocols and two channels, there are $8$ transition rates which satisfy $4$ local detailed balance relations. The current is again defined as the heat transferred from the system to the hot bath, and its fluctuation is calculated from full counting statistics. 

The meter is initially set to be in state $0$. Let the probability of the system being found in state $u (d)$ be $p_{u}^{\textrm{ini}}(p_{d}^{\textrm{ini}})$. The meter and the system are assumed to be uncorrelated initially. Thus we have
\begin{eqnarray}
\mathbf{p}_{\rm{s}}^{\textrm{ini}}&=&[p_{u}^{\textrm{ini}}, p_{d}^{\textrm{ini}}]^{\mathbb{T}},\\
\mathbf{p}_{\rm m}^{\textrm{ini}}&=&[1, 0]^{\mathbb{T}},\\
\mathbf{p}^{\textrm{ini}}&=&[p_{u}^{\textrm{ini}}, p_{d}^{\textrm{ini}},0,0]^{\mathbb{T}} = \mathcal{C}\mathbf{p}_{\rm s}^{\textrm{ini}},
\end{eqnarray}
where $\mathcal{C}=\left[ \begin{array}{cc} 1&0\\0&1\\ 0&0\\0&0 \end{array} \right]$ is the projection operator for the composite state.

Now suppose that the state of the system is measured.
If the system is found to be in state $u$, the conditional probability of the meter being found in state $0\ (1)$ is  $p_{0|u}=1-p\ (p_{1|u}=p)$;
if the system is found to be in state $d$, the conditional probability of the meter being found in state $0\ (1)$ is $p_{0|d}=q\ (p_{1|d}=1-q)$. Here $p$ and $q$ denote measurement errors. The post-measurement joint state is then given by
\begin{equation}
\mathbf{p}_{M}=[(1-p)p_{u}^{\textrm{ini}}, qp_{d}^{\textrm{ini}}, pp_{u}^{\textrm{ini}}, (1-q)p_{d}^{\textrm{ini}}]^{\mathbb{T}} = \mathcal{M}\mathbf{p}^{\textrm{ini}}= \mathcal{MC}\mathbf{p}_{\rm s}^{\textrm{ini}},
\end{equation}
where $\mathcal{M}=\left[ \begin{array}{cccc} 1-p&0&0&0\\0&q&0&0\\p&0&0&0\\0&1-q&0&0\end{array} \right]$ gives the effect of measurement.

Feedback control is performed according to the measurement outcome. If the meter is found to be in state $0$, then we change the energy gap between $u$ and $d$ to $\Delta_{0}=\Delta (1-w)$; if the meter is found to be in state $1$, we change it to $\Delta_{1}=\Delta (1+w)$.
The transition rate matrix is then given by
\begin{align}
\mathcal{R}=\mathcal{R}^{h}+\mathcal{R}^{c}=\left( 
\begin{array}{cc} \mathcal{R}_{0}^{h}&\mathbf{0}\\\mathbf{0}&\mathcal{R}_{1}^{h}\end{array} \right)+\left( 
\begin{array}{cc} \mathcal{R}_{0}^{c}&\mathbf{0}\\\mathbf{0}&\mathcal{R}_{1}^{c}\end{array} \right),
\label{r}
\end{align}
where $\mathcal{R}_{0}^{h}=\left(\begin{array}{cc} -a_h&b_h\\a_h&-b_h\end{array} \right)$ and $\mathcal{R}_{1}^{h}=\left(\begin{array}{cc} -c_h&d_h\\c_h&-d_h\end{array} \right)$, with $a_h=r^h(0u,0d)$, $b_h=r^h(0d,0u)$, $a_c=r^c(0u,0d)$, $b_c=r^c(0d,0u)$, $c_h=r^h(1u,1d)$, $d_h=r^h(1d,1u)$, $c_c=r^c(1u,1d)$ and $d_c=r^c(1d,1u)$. These 8 rates satisfy 4 local detailed balance conditions:
\begin{align}
\frac{a_h}{b_h}=e^{\beta_{h}\Delta_{0}},  \ \frac{a_c}{b_c}=e^{\beta_{c}\Delta_{0}}, \  \frac{c_h}{d_h}=e^{\beta_{h}\Delta_{1}},  \ \frac{c_c}{d_c}=e^{\beta_{c}\Delta_{1}}.
\end{align}

After the feedback control, the system evolves according to the master equation for an interaction time $\tau$:
\begin{align}
\frac{\textrm{d}\mathbf{p}}{\textrm{d}t}=\mathcal{R}\mathbf{p}.
\end{align}
At the end of one cycle, the marginal state of the system is
\begin{align}
\mathbf{p}_{\rm s}(\tau)=\mathcal{P}^{\rm s}\mathbf{p}(\tau)=\mathcal{P}^{\rm s}e^{\mathcal{R}\tau}\mathcal{MC}\mathbf{p}_{\rm s}^{\textrm{ini}}\equiv \mathcal{T}\mathbf{p}_{\rm s}^{\textrm{ini}},
\end{align}
where $\mathcal{P}^{\rm s}=\left[ \begin{array}{cccc} 1&0&1&0\\0&1&0&1\end{array} \right]$ is the projector onto the marginal state of the system. Finally, we reset the state of the meter to the default state $0$. After many cycles, the system will reach a stroboscopic steady state:
\begin{align}
\mathcal{T}\mathbf{p}_{\rm{s,sss}}^{\textrm{ini}}=\mathbf{p}_{\rm{s,sss}}^{\textrm{ini}}.
\end{align}
Thus the stroboscopic steady state is given by the eigenvector associated with eigenvalue $1$ of matrix $\mathcal{T}$. It can be calculated explicitly. As can be seen from Eq.~(\ref{r}), the sectors with $m=0$ and $1$ are completely isolated from each other, and full counting statistics can be calculated independently for each sector in a manner similar to what is done in the second example discussed in Sec. II (see Eq.~(\ref{Z})). In our simulation, we set $r^{h}(0u,0d)=3$, $r^{c}(0u,0d)=2$, $r^{h}(1u,1d)=2$, $r^{c}(1u,1d)=1$, $\beta_c=2$, $\beta_h=1$, $p=q=0$ and $\Delta=1$.

\begin{figure}
\begin{center}
\includegraphics[width=8.5cm]{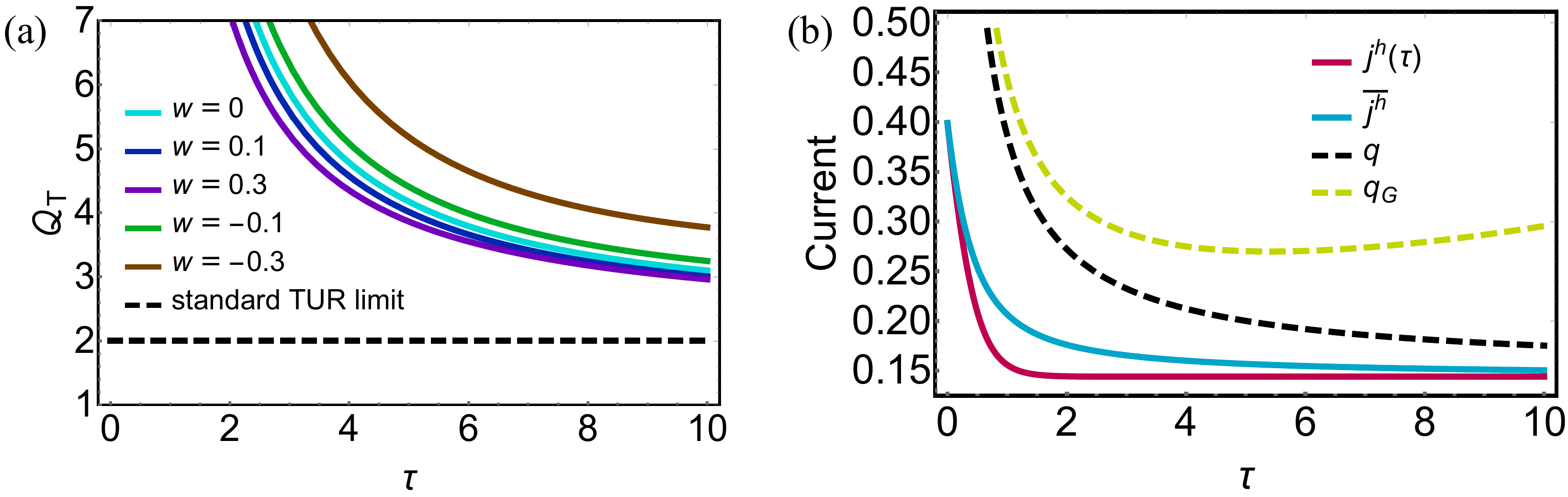}
\end{center}
\caption{(a) The solid curves show $\mathcal{Q}_{\rm T}$ for different values of the feedback control parameter $w$. The black dashed line shows the lower bound $2$ for the conventional TUR limit. It can be seen that our bound is always satisfied. (b) Dependence of the current on the period $\tau$ of the feedback cycle for the control parameter $w=0.3$. The red and blue solid curves show the final instantaneous current and the time-averaged one, respectively. The black and green dashed curves represent the current bounds associated with the conventional TUR and the GTUR. Both the final current and the time-averaged one decrease with $\tau$ as discussed in the main text.}
\label{figS3}
\end{figure}

As shown in Fig.~\ref{figS3}(a), for different control parameters $w$, our bound always holds. We note that in this model, the current monotonically decreases, implying that $\mathcal{Q}_{\rm C}<\mathcal{Q}_{\rm T}$. Although for our chosen parameter set $\mathcal{Q}_{\rm C}$ does not surpass the bound $2$, there could be a possible violation for a wider parameter range. In Fig.~\ref{figS3}(b), we show the current as a function of the time period $\tau$ for the feedback control parameter $w=0.3$. It can be seen that both the final current and the time-averaged one decrease with $\tau$ and that they are bounded from above by $q$ and $q_{G}$. As stated in the main text, the decreasing current implies a possible violation in the conventional TUR. While the conventional TUR is not violated for the parameter range chosen here, it might break down for a wider range of the parameters.

\section{Some details in the derivation of the main result (6)}
First, we derive Eq.~(14) in the main text.
\begin{align}
&F(\theta)=-\bigg\langle \frac{\partial^2}{\partial \theta^2}\ln \mathcal{P}_{\theta}[\omega] \bigg\rangle_{\theta}\nonumber\\
=&\int_{0}^{T}dt\sum_{x\ne y,\nu}\bigg[-P_{\theta}(x;t)\frac{\partial^2}{\partial \theta^2}r^{\nu}_{\theta}(x,y;t)\nonumber\\
&+P_{\theta}(x;t)r^{\nu}_{\theta}(x,y;t)\frac{\partial^2}{\partial \theta^2}\ln r^{\nu}_{\theta}(x,y;t)\bigg]\nonumber\\
=&\int_{0}^{T}dt\sum_{x\ne y,\nu}P_{\theta}(x;t)r^{\nu}_{\theta}(x,y;t)\bigg(\frac{\partial}{\partial \theta}\ln r^{\nu}_{\theta}(x,y;t)\bigg)^2.
\end{align}

Then, we prove that Eqs.~(16) and (17) always have solutions. We first note that they take the following forms:
\begin{align}
&AX^2+BY^2=\frac{1}{2}(A-B)\ln\frac{A}{B},\label{EPequation}\\
&AX-BY=A-B.\label{currentequation}
\end{align}
We use Eq.~(\ref{currentequation}) to eliminate $Y$ in Eq.~(\ref{EPequation}):
\begin{equation}
A(A+B)X^2-2A(A-B)X+(A-B)^2-\frac{1}{2}(A-B)B\ln\frac{A}{B}=0.
\end{equation}
The discriminant of this equation can be shown to be non-negative:
\begin{align}
\Delta&=2AB(A+B)(A-B)\ln\frac{A}{B}-4AB(A-B)^2\nonumber\\
&\ge 4AB(A+B)(A-B)^2-4AB(A-B)^2=0,
\end{align}
where the inequality $(a-b)\ln\frac{a}{b}\ge\frac{2(a-b)^2}{a+b}$ for $a,b>0$ is used. Thus the equations always have a solution. Explicitly, it can be written as
\begin{align}
\alpha^\nu_{xy}=X=\j_{xy}^\nu\pm\j_{xy}^\nu\sqrt{\frac{1-\j_{xy}^\nu}{1+\j_{xy}^\nu}\bigg(\frac{F_{xy}^\nu}{2\j_{xy}^\nu}-1\bigg)},
\label{solution}
\end{align}
where $\j_{xy}^\nu(t)\equiv \frac{j^\nu(x,y;t)}{t^\nu(x,y;t)}$ is the current normalized by the traffic rate $t^\nu(x,y;t)\equiv P(x;t)r^\nu(x,y)+P(y;t)r^\nu(y,x)$, and $F_{xy}^\nu(t)\equiv \ln\frac{P(x;t)r^\nu(x,y)}{P(y;t)r^\nu(y,x)}$ is the thermodynamic force at time $t$. In Eq.~(\ref{solution}), the time dependence is omitted for simplicity.

To prove Eq.~(18), we have to prove another important relation. For small $\theta$, we expand $\mathcal{R}_{\theta}(t)$ up to the first order in $\theta$:
\begin{align}
\mathcal{R}_{\theta}(t)=\mathcal{R}+\theta \mathcal{R}_{1}(t)+O(\theta^2).
\end{align}
We use Eq.~(17) to show that $\mathcal{R}_{1}(t)$ satisfies
\begin{align}
[\mathcal{R}_{1}(t)\mathbf{P}(t)]_{y}&=\sum_{x:x\ne y,\nu}\big[K^\nu_{xy}\alpha^\nu_{xy}-K^\nu_{yx}\alpha^\nu_{yx}\big]\nonumber\\
&=\sum_{x:x\ne y,\nu}(K^\nu_{xy}-K^\nu_{yx})=[\mathcal{R}\mathbf{P}(t)]_{y}.\label{R1P}
\end{align}
We use this relation to prove Eq.~(18). Instead of $\mathbf{P}_{\theta}(t)$, we consider its transformed quantity:
\begin{align}
\tilde{\mathbf{P}}_{\theta}(t)\equiv e^{-\mathcal{R}t}\mathbf{P}_{\theta}(t).
\end{align}
The time derivative of this quantity can be calculated as
\begin{align}
\dot{\tilde{\mathbf{P}}}_{\theta}(t)&=-\mathcal{R}\tilde{\mathbf{P}}_{\theta}(t)+e^{-\mathcal{R}t}\dot{\mathbf{P}}_{\theta}(t)\nonumber\\
&=-\mathcal{R}\tilde{\mathbf{P}}_{\theta}(t)+e^{-\mathcal{R}t}(\mathcal{R}+\theta \mathcal{R}_{1}(t))\mathbf{P}_{\theta}(t)\nonumber\\
&=\theta e^{-\mathcal{R}t}\mathcal{R}_{1}(t)e^{\mathcal{R}t}\tilde{\mathbf{P}}_{\theta}(t).
\end{align}
Integrating this equation from $0$ to $t$ gives
\begin{align}
\tilde{\mathbf{P}}_{\theta}(t)&=\mathbf{P}(0)+\theta\int_{0}^{t}ds e^{-\mathcal{R}s}\mathcal{R}_{1}(s)\mathbf{P}_{\theta}(s)\nonumber\\
&=\mathbf{P}(0)+\theta\int_{0}^{t}ds e^{-\mathcal{R}s}\mathcal{R}_{1}(s)\mathbf{P}(s)+O(\theta^2).
\end{align}
We use Eq.~\eqref{R1P} to obtain
\begin{align}
\tilde{\mathbf{P}}_{\theta}(t)&=\mathbf{P}(0)+\theta\int_{0}^{t}ds e^{-\mathcal{R}s}\mathcal{R}\mathbf{P}(s)+O(\theta^2)\nonumber\\
&=\mathbf{P}(0)+\theta t \mathcal{R}\mathbf{P}(0)+O(\theta^2).
\end{align}
We thus arrive at
 \begin{align}
 \mathbf{P}_\theta(t)=  \mathbf{P}(t)+\theta t \dot{\mathbf{P}}(t)+O(\theta^2).
 \end{align}
 \
 
 \section{Alternative Derivation of the main result (6) from the large deviation theory}
In this alternative derivation, we employ the level 2.5 large deviation theory which is a slight modification of that used for periodically driven dynamics \cite{Barato2018} and finite-time TUR \cite{Horowitz2017}. As in Ref.~\cite{Horowitz2017}, we consider a large ensemble of $N$ copies of the dynamics. The joint probability distribution of empirical states $\tilde{P}(x;t)$ and empirical currents $\tilde{j}^{\nu}(x,y;t)$, which deviate from their typical values, should be exponentially small for large $N$ as
\begin{align}
P(\tilde{P}(t),\tilde{j}(t))\asymp e^{-NI(\tilde{P}(t),\tilde{j}(t))},
\end{align}
where $\asymp$ denotes the asymptotic logarithmic equivalence \cite{Horowitz2017}. The rate function is
\begin{align}
I(\tilde{P}(t),\tilde{j}(t))=\int_{0}^{T}dtL(\tilde{P}(t),\tilde{j}(t))+D_{\textrm{KL}}(\tilde{\mathbf{P}}(0)||\mathbf{P}(0)),
\label{rate}
\end{align}
where
\begin{align}
D_{\textrm{KL}}(\tilde{\mathbf{P}}(0)||\mathbf{P}(0))\equiv\sum_{x}\tilde{P}(x;0)\ln\frac{\tilde{P}(x;0)}{P(x;0)}
\end{align}
is the Kullback-Leibler divergence of the empirical initial distribution $\tilde{\mathbf{P}}(0)$ from the typical one $\mathbf{P}(0)$, and
\begin{align}
L(\tilde{P}(t),\tilde{j}(t))=\sum_{x>y,\nu}\Psi(\tilde{j}^{\nu}(x,y;t),j^{\nu}_{\tilde{P}}(x,y;t),a^{\nu}_{\tilde{P}}(x,y;t)), \label{LPj}
\end{align}
with
\begin{equation}
\Psi(j,\bar{j},a)\equiv j\bigg(\textrm{arsinh}\frac{j}{a}-\textrm{arsinh}\frac{\bar{j}}{a}\bigg)-\sqrt{a^2+j^2}-\sqrt{a^2+\bar{j}^2}.
\end{equation}
Here the quantities in Eq.~\eqref{LPj} are defined as
\begin{align}
&\tilde{j}^{\nu}(x,y;t)\equiv\tilde{P}(x;t)\tilde{r}^{\nu}(x,y;t)-\tilde{P}(y;t)\tilde{r}^{\nu}(y,x;t),\\
&j^{\nu}_{\tilde{P}}(x,y;t)\equiv\tilde{P}(x;t)r^{\nu}(x,y)-\tilde{P}(y;t)r^{\nu}(y,x),\\
&a^{\nu}_{\tilde{P}}(x,y;t)\equiv 2\sqrt{\tilde{P}(x;t)\tilde{P}(y;t)r^{\nu}(x,y)r^{\nu}(y,x)}.
\end{align}

It has been shown \cite{Horowitz2017,Barato2018} that the rate function in Eq.~(\ref{rate}) has a parabolic upper bound as 
\begin{widetext}
\begin{align}
I(\tilde{P}(t),\tilde{j}(t))\le \int_{0}^{T}dt\sum_{x>y,\nu}\bigg[\frac{\tilde{j}^{\nu}(x,y;t)-j^{\nu}_{\tilde{P}}(x,y;t)}{2j^{\nu}_{\tilde{P}}(x,y;t)}\bigg]^2\sigma^{\nu}_{\tilde{P}}(x,y;t)+D_{\textrm{KL}}(\tilde{\mathbf{P}}(0)||\mathbf{P}(0)),
\label{bound}
\end{align}
\end{widetext}
where
\begin{align}
\sigma^{\nu}_{\tilde{P}}(x,y;t)\equiv j^{\nu}_{\tilde{P}}(x,y;t)\ln\frac{\tilde{P}(x;t)r^{\nu}(x,y)}{\tilde{P}(y;t)r^{\nu}(y,x)}.
\end{align}

The empirical time-integrated current is given by
\begin{align}
J_d\equiv\int_{0}^{T}dt\sum_{x\ne y,\nu}\tilde{j}^{\nu}(x,y;t)d^{\nu}(x,y).
\label{Jd}
\end{align}
To obtain the rate function of this empirical accumulated current $J_d$, the contraction principle should be employed under two constraints \cite{Horowitz2017,Barato2018}. One is the normalization condition for the empirical distribution $\sum_{x}\tilde{P}(x;t)=1$. The other is a master equation obeyed by the empirical transition rates and the empirical states, i.e., $\dot{\tilde{P}}(x;t)=\sum_{y:y\ne x,\nu}\tilde{j}^{\nu}(y,x;t)$.

We let the empirical quantities be the same as the parametrized ones in the derivation via the Cram\'e-Rao inequality (see Eqs.~(16) and (18) in the main text). They are written as
\begin{align}
&\tilde{r}^{\nu}(x,y;t)=r^{\nu}(x,y)e^{\theta\alpha_{xy}^{\nu}(t)},\\
&\tilde{P}(x;t)=P(x;t)+\theta t \dot{P}(x;t)+O(\theta^2).
\end{align}
This choice ensures that the above two constraints are satisfied. It can be shown that
\begin{align}
&\tilde{j}^{\nu}(x,y;t)=(1+\theta)j^{\nu}(x,y;t)+\theta t \dot{j}^{\nu}(x,y;t)+O(\theta^2),\\
&j^{\nu}_{\tilde{P}}(x,y;t)=j^{\nu}(x,y;t)+\theta t \dot{j}^{\nu}(x,y;t)+O(\theta^2),\\
&\sigma^{\nu}_{\tilde{P}}(x,y;t)=\sigma^{\nu}(x,y;t)+O(\theta),
\end{align}
where
\begin{align}
\sigma^{\nu}(x,y;t)\equiv j^{\nu}(x,y;t)\ln\frac{P(x;t)r^{\nu}(x,y)}{P(y;t)r^{\nu}(y,x)}
\end{align}
is the typical entropy production rate for the transition between states $x$ and $y$ via channel $\nu$. The relative entropy $D_{\textrm{KL}}(\tilde{\mathbf{P}}(0)||\mathbf{P}(0))$  then vanishes. From the definition of the empirical accumulated current $J_d$ Eq.~(\ref{Jd}), $\theta$ is determined as
\begin{align}
\theta=\frac{J_d-\langle J \rangle}{Tj(T)}.
\end{align}

Finally, by contracting Eq.~(\ref{bound}), the bound on the rate function of $J_d$  is given by
\begin{align}
I(J_d)\le \frac{1}{4}\bigg[\frac{J_d-\langle J \rangle}{Tj(T)}\bigg]^2\sigma+O(\theta^3).
\end{align}
Hence,
\begin{align}
\textrm{Var}[J]=\frac{1}{I''(\langle J \rangle)}\ge 2\frac{(Tj(T))^2}{\sigma},
\end{align}
which is the desired bound, i.e., Eq.~(6) in the main text.

\section{Derivation of the discrete-time TUR (11)}
The parametrized path probability is given as
\begin{align}
\mathcal{P}_{\theta}[\omega]=P_{\theta}(x_0)\prod_{i=1}^{n}A^{\nu_i}_{\theta}(x_{i}|x_{i-1};t_{i-1}),
\end{align}
where the parametrized transition probability $A^{\nu_i}_{\theta}(x_{i}|x_{i-1};t_{i-1})$ depends on the time $t_{i-1}$ and the time step $\Delta t$. By definition, the Fisher information involves an off-diagonal part and a diagonal one:
\begin{align}
F(\theta)=F_{\textrm{off-dia}}(\theta)+F_{\textrm{dia}}(\theta),
\end{align}
where
\begin{align}
&F_{\textrm{off-dia}}(\theta)\nonumber\\
&\equiv\sum_{i=1}^{n}\sum_{x\ne y,\nu}P_{\theta}(x,t_{i-1})A_{\theta}^{\nu}(y|x;t_{i-1})\bigg[\frac{\partial}{\partial\theta}\ln A_{\theta}^{\nu}(y|x;t_{i-1})\bigg]^2, \label{Foffdia}\\
&F_{\textrm{dia}}(\theta)\nonumber\\
&\equiv\sum_{i=1}^{n}\sum_{x}P_{\theta}(x,t_{i-1})A_{\theta}(x|x;t_{i-1})\bigg[\frac{\partial}{\partial\theta}\ln A_{\theta}(x|x;t_{i-1})\bigg]^2.\label{Fdia}
\end{align}
The total EP can be derived from a change in the Shannon entropy of the system per unit step. For a step between $t_{i-1}$ and $t_i$, the EP in the heat reservoirs is given by
\begin{align}
\Delta\sigma_{B_i}=\sum_{x,y,\nu}P(x,t_{i-1})A^{\nu}(y|x)\ln\frac{A^{\nu}(y|x)}{A^{\nu}(x|y)}.
\end{align}
The Shannon entropy change in the system is
\begin{equation}
\Delta\sigma_{S_i}=\sum_{x}\big[-P(x,t_i)\ln P(x,t_i)+P(x,t_{i-1})\ln P(x,t_{i-1})\big].
\end{equation}
By defining \cite{Morimoto1963,Lee2018}, we have
\begin{align}
&\Delta P(x,t_{i-1})\equiv P(x,t_i)-P(x,t_{i-1}),\\
&k^{\nu}(x|y)\equiv A^{\nu}(x|y)-\delta_{xy},
\end{align}
where $\delta_{xy}$ is the Kronecker delta, and a change in the Shannon entropy in one step can be calculated as
\begin{align}
\Delta\sigma_{S_i}=\sum_{x,y,\nu}P(x,t_{i-1})A^{\nu}(y|x)\ln\frac{P(x,t_{i-1})}{P(y,t_i)}.
\end{align}
Therefore, the total EP in one step reads
\begin{align}
\Delta\sigma_i=\sum_{x,y,\nu}P(x,t_{i-1})A^{\nu}(y|x)\ln\frac{P(x,t_{i-1})A^{\nu}(y|x)}{P(y,t_i)A^{\nu}(x|y)}.
\end{align}
For $n$ steps, the total EP should be
\begin{align}
\sigma=\sum_{i=1}^{n}\sum_{x,y,\nu}P(x,t_{i-1})A^{\nu}(y|x)\ln\frac{P(x,t_{i-1})A^{\nu}(y|x)}{P(y,t_i)A^{\nu}(x|y)},\label{discreteEP}
\end{align}
As we can see, the probabilities in the logarithm are not equal-time, which is different from the continuous-time case. This feature prevents us from rewriting Eq.~\eqref{discreteEP} into a form of $(A-B)\ln\frac{A}{B}$ as in the continuous-time one. However, the key equation (16) in the main text requires such a form. We have to modify the total EP into the tilde EP as
\begin{align}
\tilde{\sigma}&\equiv\sigma+\sum_{i=1}^{n}D_{\textrm{KL}}(\mathbf{P}(t_i)||\mathbf{P}(t_{i-1}))\nonumber\\
&=\sum_{i=1}^{n}\sum_{x,y,\nu}P(x,t_{i-1})A^{\nu}(y|x)\ln\frac{P(x,t_{i-1})A^{\nu}(y|x)}{P(y,t_{i-1})A^{\nu}(x|y)}.\label{tildeEP}
\end{align}

By choosing the parametrized transition probabilities as 
\begin{align}
A_{\theta}^{\nu}(y|x;t_{i-1})=A^{\nu}(y|x)e^{\theta\alpha^{\nu}(y|x;t_{i-1})},
\end{align}
similar assumptions as in Eqs.~(16) and (17) can be made as well:
\begin{align}
&K_{xy}^{\nu}(\alpha_{xy}^{\nu})^{2}+K_{yx}^{\nu}(\alpha_{yx}^{\nu})^{2}=\frac{1}{2}(K_{xy}^{\nu}-K_{yx}^{\nu})\ln\frac{K_{xy}^{\nu}}{K_{yx}^{\nu}}, \label{ep}\\
&K_{xy}^{\nu}\alpha_{xy}^{\nu}-K_{yx}^{\nu}\alpha_{yx}^{\nu}=K_{xy}^{\nu}-K_{yx}^{\nu} \label{j},
\end{align}
where $K_{xy}^{\nu}(t_{i-1})\equiv P(x,t_{i-1})A^{\nu}(y|x)$ and $\alpha_{xy}^{\nu}(t_{i-1})\equiv \alpha^{\nu}(y|x;t_{i-1})$. It is clear that Eq.~\eqref{ep} gives
\begin{align}
F_{\textrm{off-dia}}(0)=\frac{\tilde{\sigma}}{2}.
\end{align}
The diagonal term of the Fisher information is shown to be bounded from above as
\begin{align}
F_{\textrm{dia}}(0)&=\sum_{i=1}^{n}\sum_{x}P(x)\frac{[\sum_{y:y\ne x,\nu}A^{\nu}(y|x)\alpha^{\nu}_{xy}(t_{i-1})]^2}{1-\sum_{y:y\ne x,\nu}A^{\nu}(y|x)}\nonumber\\
&\le \sum_{i=1}^{n}\sum_{x\ne y,\nu}P(x)A^{\nu}(y|x)(\alpha^{\nu}_{xy}(t_{i-1}))^2\frac{1-A(x|x)}{A(x|x)}\nonumber\\
&\le \sum_{i=1}^{n}\frac{1-\min_{x}A(x|x)}{\min_{x}A(x|x)}\sum_{x\ne y,\nu}P(x)A^{\nu}(y|x)(\alpha^{\nu}_{xy}(t_{i-1}))^2\nonumber\\
&\equiv \frac{1-a}{a}\frac{\tilde{\sigma}}{2},
\end{align}
where Titu's lemma $\frac{(\sum_{i}u_i)^2}{\sum_{i}v_i}\le \sum_{i}\frac{u_i^2}{v_i}$ has been used in the first inequality, and in the second inequality we take the minimal diagonal entry of $\mathcal{A}$, namely the minimal staying probability, denoted as $a\equiv \min_{x}A(x|x)$ which satisfies $0<a<1$. Finally, we obtain an upper bound on the Fisher information:
\begin{align}
F(0)\le \frac{1}{a}\frac{\tilde{\sigma}}{2}. \label{Fbound}
\end{align}

For the current, similarly to the continuous-time case, we expand the parametrized transition probability matrix up to the first order in $\theta$ as
\begin{align}
\mathcal{A}_{\theta}(t_{i-1})=\mathcal{A}+\theta \mathcal{A}_{1}(t_{i-1})+\mathcal{O}(\theta^2).
\end{align}
It can be shown that $\mathcal{A}_{1}(t_{i-1})$ satisfies the following important property similar to Eq.~\eqref{R1P}:
\begin{align}
[\mathcal{A}_{1}(t_{i-1})\mathbf{P}(t_{i-1})]_{y}&=\sum_{x,\nu}P(x,t_{i-1})A^{\nu}(y|x)\alpha^{\nu}_{xy}(t_{i-1})\nonumber\\
&=\sum_{x:x\ne y,\nu}(K_{xy}^{\nu}\alpha_{xy}^{\nu}-K_{yx}^{\nu}\alpha_{yx}^{\nu})\nonumber\\
&=\sum_{x:x\ne y,\nu}(K_{xy}^{\nu}-K_{yx}^{\nu})\nonumber\\
&=[\mathbf{P}(t_{i})-\mathbf{P}(t_{i-1})]_y, \label{AP}
\end{align}
where Eq.~(\ref{j}) is used in the third equality. From Eq.~\eqref{AP}, the parametrized state probabilities can be calculated by iteration as
\begin{align}
\mathbf{P}_{\theta}(t_{i})&=\mathcal{A}_{\theta}(t_{i-1})\mathbf{P}_{\theta}(t_{i-1})\nonumber\\
&=[\mathcal{A}+\theta \mathcal{A}_{1}(t_{i-1})]\mathbf{P}_{\theta}(t_{i-1})+\mathcal{O}(\theta^2)\nonumber\\
&=[\mathcal{A}+\theta \mathcal{A}_{1}(t_{i-1})][\mathcal{A}+\theta \mathcal{A}_{1}(t_{i-2})]\mathbf{P}_{\theta}(t_{i-2})+\mathcal{O}(\theta^2)\nonumber\\
&\dots\nonumber\\
&=[\mathcal{A}+\theta \mathcal{A}_{1}(t_{i-1})]\cdots[\mathcal{A}+\theta \mathcal{A}_{1}(0)]\mathbf{P}_{\theta}(0)+\mathcal{O}(\theta^2)\nonumber\\
&=\{\mathcal{A}^{i}+\theta [\mathcal{A}_{1}(t_{i-1})\mathcal{A}^{i-1}+\cdots +\mathcal{A}^{i-1}\mathcal{A}_{1}(0)]\}\mathbf{P}(0)\nonumber\\
&\quad+\mathcal{O}(\theta^2)\nonumber\\
&=\mathbf{P}(t_{i})+i\theta[\mathbf{P}(t_{i})-\mathbf{P}(t_{i-1})]+\mathcal{O}(\theta^2).
\end{align}
Hence
\begin{align}
\frac{\partial}{\partial\theta}P_{\theta}(x,t_{i})\bigg|_{\theta=0}=i[P(x,t_{i})-P(x,t_{i-1})].
\end{align}
It is now possible to calculate $\psi'(0)$ in the Cram\'er-Rao inequality:
\begin{widetext}
\begin{align}
\psi'(0)&=\sum_{i=1}^{n}\sum_{x\ne y,\nu}\bigg[P_{\theta}(x,t_{i-1})\frac{\partial}{\partial\theta}A_{\theta}^{\nu}(y|x;t_{i-1})+A_{\theta}^{\nu}(y|x;t_{i-1})\frac{\partial}{\partial\theta}P_{\theta}(x,t_{i-1})\bigg]d^{\nu}(y|x)\bigg|_{\theta=0}\nonumber\\
&=\sum_{i=1}^{n}\sum_{x\ne y,\nu}\big\{P(x,t_{i-1})+(i-1)[P(x,t_{i-1})-P(x,t_{i-2})]\big\}A^{\nu}(y|x)d^{\nu}(y|x)\nonumber\\
&=\sum_{i=1}^{n}\sum_{x\ne y,\nu}iP(x,t_{i-1})A^{\nu}(y|x)d^{\nu}(y|x)-\sum_{i=1}^{n-1}\sum_{x\ne y,\nu}iP(x,t_{i-1})A^{\nu}(y|x)d^{\nu}(y|x)\nonumber\\
&=n\sum_{x\ne y,\nu}P(x,t_{n-1})A^{\nu}(y|x)d^{\nu}(y|x)\equiv nj(t_{n-1}),\label{discretepsi}
\end{align}
\end{widetext}
where the current at the final step is defined as
\begin{align}
j(t_{n-1})\equiv \sum_{x\ne y,\nu}P(x,t_{n-1})A^{\nu}(y|x)d^{\nu}(y|x).
\end{align}
Combining inequality~\eqref{Fbound} and Eq.~\eqref{discretepsi}, we obtain the desired TUR (11).

\section{Relations among the four TURs}
\begin{figure}
\begin{center}
\includegraphics[width=8.5cm]{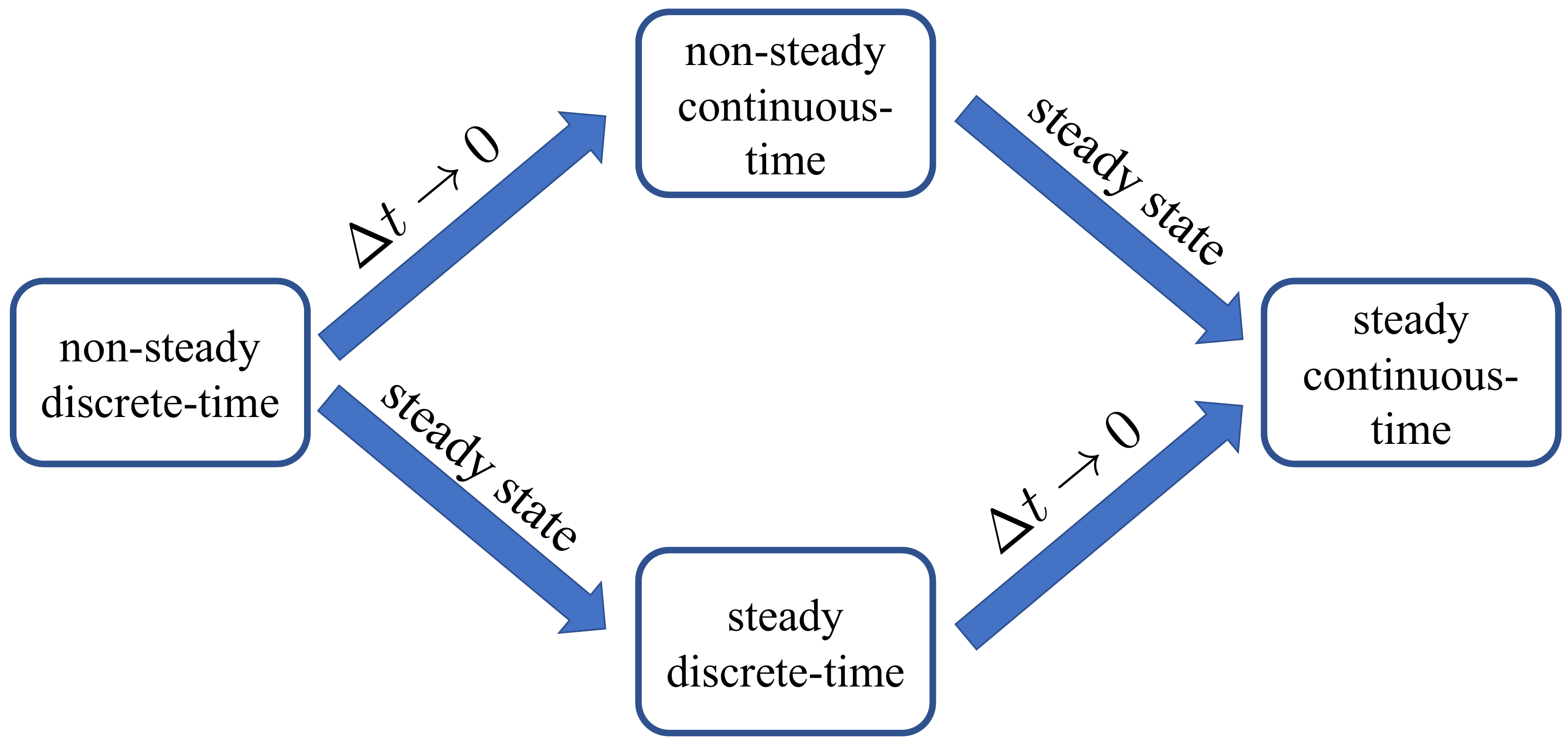}
\end{center}
\caption{Hierarchical relations among the four TURs. The non-steady-state discrete-time TUR (11) reduces to the non-steady-state continuous-time TUR (6) in the continuous-time limit, namely in the limit where the time step vanishes: $\Delta t\to 0$. The non-steady state discrete-time TUR reduces to the steady-state discrete-time  TUR \eqref{SSDT} for the steady initial state. The non-steady continuous-time TUR  and the steady-state discrete-time TUR reduce to the original steady-state continuous-time TUR, namely the conventional TUR (1), for steady initial states and the continuous-time limit, respectively.}
\label{TURs}
\end{figure}

There are hierarchical relations among all the bounds obtained in Fig.~\ref{TURs} in which the non-steady-state discrete-time TUR (11) is the most general one. All other three TURs can be deduced from it by considering proper limits, and the conventional TUR (1) can be reduced from any of the other bounds.

Specifically, we prove that the TUR (6) for continuous-time dynamics can be reduced from the discrete-time TUR (11) by taking the continuous-time limit, namely $\Delta t\to0$. In this limit, we denote $P(x,t_{i})=P(x,t_{i-1})+q(x,t_{i-1})\Delta t$. According to the normalization condition, we have $\sum_{x}q(x,t_{i-1})=0$, the Kullback-Leibler (KL) divergence in the tilde EP \eqref{tildeEP} is $ \mathcal{O}(\Delta t^2)$:
\begin{align}
D_{\textrm{KL}}(\mathbf{P}(t_i)||\mathbf{P}(t_{i-1}))&\equiv\sum_{x}P(x,t_{i})\ln\frac{P(x,t_{i})}{P(x,t_{i-1})}\nonumber\\
&=\sum_{x}q(x,t_{i-1})\Delta t+\sum_{x}\frac{q(x,t_{i-1})^2}{P(x,t_{i-1})}\Delta t^2 \nonumber\\
&=\mathcal{O}(\Delta t^2).
\end{align}
The sum of the KL divergences is thus $\mathcal{O}(\Delta t)$:
\begin{align}
\sum_{i=1}^{n}D_{\textrm{KL}}(\mathbf{P}(t_i)||\mathbf{P}(t_{i-1}))=n\mathcal{O}(\Delta t^2)=\mathcal{O}(\Delta t).
\end{align}
Because the total EP $\sigma$ is $\mathcal{O}(1)$, the tilde EP $\tilde{\sigma}$ is approximately the total EP $\sigma$. The transition probabilities can be expressed in terms of transition rates and time step as
\begin{align}
&A^{\nu}(y|x)=r^{\nu}(x,y)\Delta t,\\
&A(x|x)=1-\sum_{y:y\ne x,\nu}A^{\nu}(y|x)=1-\sum_{y:y\ne x,\nu}r^{\nu}(x,y)\Delta t.
\end{align}
The minimal staying probability is hence approximately given by $a=1-\mathcal{O}(\Delta t)$. Eventually, we obtain
\begin{align}
\frac{\tilde{\sigma}}{a}\to\sigma.
\end{align}

The steady-state discrete-time TUR can simply be obtained from the bound (11) for the steady initial state as:
\begin{align}
\frac{\textrm{Var}[J]}{\langle J \rangle^2}\frac{\sigma}{a}\ge 2.\label{SSDT}
\end{align}
There exists a generalized TUR-like TUR valid for discrete-time Markov chains for an NESS \cite{Proesmans2017}:
\begin{align}
\frac{\textrm{Var}[J]}{\langle J\rangle^2}(e^{\Delta\sigma_i}-1)\ge 2.\label{DTUR}
\end{align}
Compared with this result, our steady-state discrete-time TUR \eqref{SSDT} is linear in the total EP and we have a new linear bound (11) valid for arbitrary initial states. Thus we have greatly generalized the existing results in the discrete-time regime.

\end{document}